# Pinning in a Porous Bi2223

K. Yu. Terent'ev, D. M. Gokhfeld, S. I. Popkov, K. A. Shaykhutdinov, and M. I. Petrov

*Kirensky Institute of Physics, Siberian Branch of the Russian Academy of Sciences,
Akademgorodok 50 38, Krasnoyarsk, 660036 Russia*

**Abstract** The current–voltage characteristics of a porous superconductor $Bi_2Sr_2Ca_2Cu_3O_x$ (Bi2223) have been measured at temperature range from 10 to 90 K. The experimental dependences have been analyzed within the model allowing for pinning by clusters of a normal phase with fractal boundaries, as well as the model taking into account phase transformations of vortex matter. It has been found that the electrical resistance of the superconductor material significantly increases at temperatures of 60–70 K over the entire range of magnetic fields under consideration without changing in the sign of the curvature of the dependence $R(I)$. The melting of the vortex structure occurs at these temperatures. It has been assumed that this behavior is associated with the specific feature of the pinning in a highly porous high-temperature superconductor, which lies in the fractal distribution of pinning centers in a wide range of self-similarity scales.

1. INTRODUCTION

A magnetic field penetrates into type II superconductors in the form of Abrikosov vortices. When transport current flows through such a material, the Abrikosov vortices under the action of the Lorentz force come into the motion accompanied by the appearance of a finite electrical resistance. Vortices are pinned on any inhomogeneities which exist even in high-temperature superconductor single crystals (as a rule, in the form of oxygen disordering) [1].

An additional contribution to pinning is observed in polycrystalline superconductors due to the presence of boundaries between crystallites. The superconducting properties of grain boundaries are reduced. During the flowing of the transport current, the suppression of superconductivity of the material, i.e., the growth of a cluster of the normal phase state begins to occur exactly with the grain boundaries. As was shown in [2,3], the cluster of the normal phase can be considered as a fractal that changes its own dimension with increasing temperature and transport current. It should be noted that superconductors with a high porosity can be successfully described in the framework of the theory of a fractal cluster of the normal phase in a superconducting matrix [4].

A random pinning has a profound effect on the lattice of Abrikosov vortices [5]. Consequently, the specific features of the structure of a porous superconductor should lead to a change in its parameters. At a high concentration of defects in the crystal, the lattice of Abrikosov vortices transforms into a glassy state. A number of authors [3,6–8] have described a vortex glass phase (similar to the spin glass phase) with a broken translational symmetry of the Abrikosov vortex lattice and a nonuniform spatial density of vortices in real imperfect polycrystals under the conditions of low temperatures and penetrating magnetic fields. The spin glass phase was considered in [9,10]. In [3], the author analyzed two types of pinning: small scale pinning (provided by inclusions, vacancies, etc.) and large scale pinning (spatially separated grains, cracks, macroscopic inclusions, etc.). As would be expected, these types of pinning should affect the transport properties of the material in different manners. The small scale pinning destroys the long order in the Abrikosov lattice, and, as a consequence, the Abrikosov lattice transforms into the glassy state. The large scale pinning, according to the author's opinion [3], affects the macroscopic regime of the vortex flow, thus leaving the vortex lattice undisturbed.

The majority of researchers have paid attention to the study of the phase diagram of Abrikosov vortices in textured polycrystals [11–16] with a predominant small scale pinning. In these works, the authors calculated the temperature of the transition to the vortex glass state $T_g$. In practice, this temperature has been found as the boundary of the change in the sign of the curvature of the scaling dependences $R(I)$ plotted in logarithmic coordinates and obtained at different temperatures. The observed change in the sign of the curvature of these dependences indicates that a phase transition occurs in the Abrikosov vortex lattice.

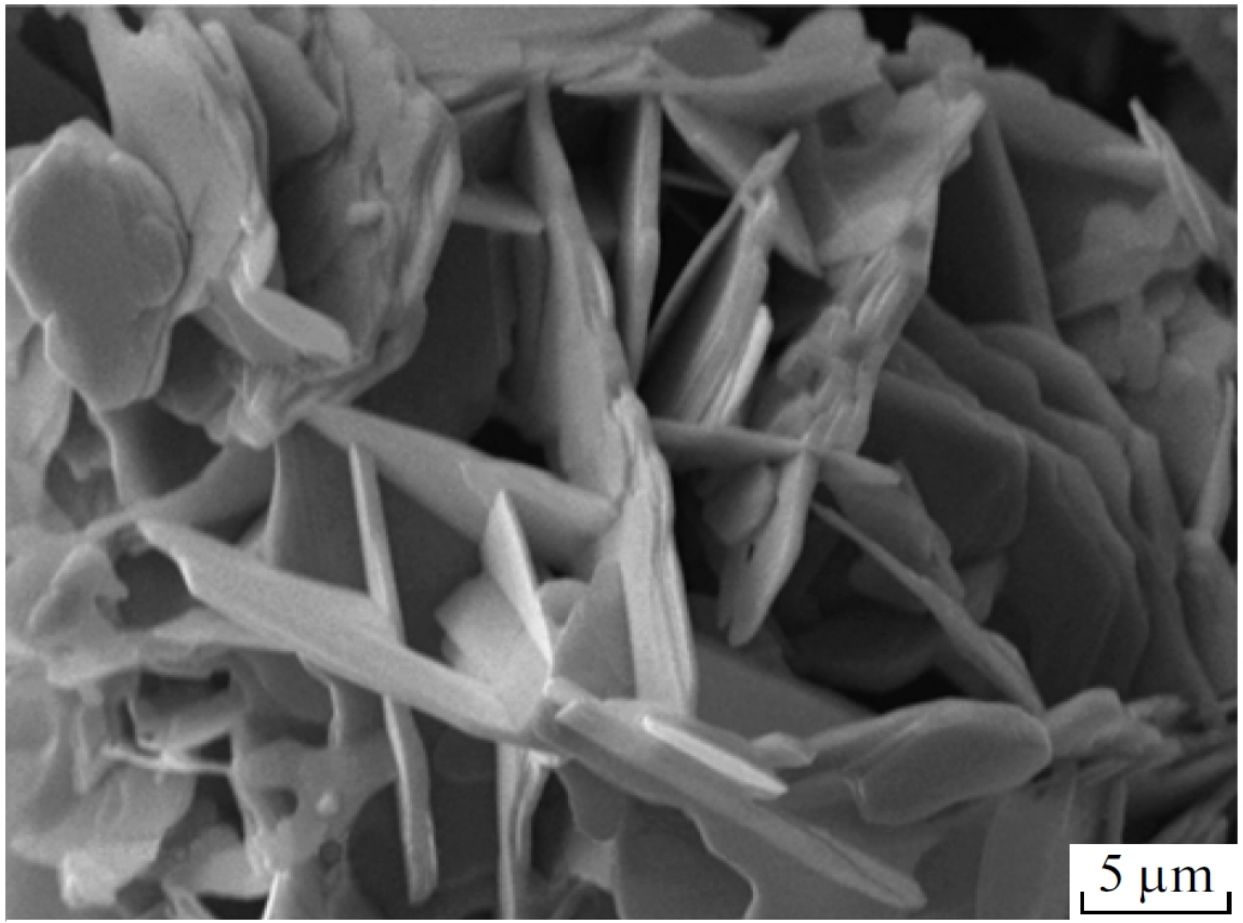

**Fig. 1.** Microstructure of the porous Bi2223 superconductor.

In a highly porous high-temperature superconductor, as expected, there exist, in a full measure, both large scale and small scale pinnings owing to the fractality of grain boundaries over a wide range of self-similarity scales. The high porosity and disordering can be favorable for a pinning and an increase in the critical current density in such materials. Indeed, in highly porous high-temperature superconductor materials, the diamagnetic response per unit mass is almost two times higher than that of the polycrystalline sample with the same composition [17]. Such material can be prepared using different methods [18–20]. A high-temperature superconductor with a porous structure that represents an aggregate of large crystallites (~20 μm) was studied earlier in [17,18]. A porous high-temperature superconductor material can be considered as an aggregate of single crystals that are spatially separated and touch each other predominantly through their faces.

In this work, we have synthesized porous samples of the Bi2223 composition and measured their current–voltage characteristics over a wide range of temperatures with the aim of investigating the specific features of the pinning in this material. The obtained results have been analyzed in the framework of the model proposed in [2,3]. We have also performed a search for phase transitions of the vortex structure [21].

2. SAMPLES AND EXPERIMENTAL TECHNIQUE

The synthesis technology used for the preparation of the porous high-temperature superconductor was described in [18]. The density of the synthesized material was equal to 1.55 g/cm3 (which amounted to 26% of the theoretical density of the Bi2223 superconductor). The scanning electron microscopy (SEM) images made it possible to examine the internal structure of the material [4]: the Bi2223 crystallites are chaotically arranged; the average size of the crystallites is ~1–2 μm along the *c* axis and ~20–30 μm in the *ab* plane; and between individual crystallites or their aggregates, there are pores with a diameter of ~10 μm and smaller (Fig. 1). The superconducting state is achieved at the temperature $T_c$ = 105 K [4].

The current–voltage characteristics were measured using the standard four point probe method in magnetic fields of 0–80 Oe and in the temperature range from 10 to 90 K. The coincidence of the forward and reverse current–voltage characteristics during the scanning of the current indicated the absence of selfheating. Figure 2 shows the current–voltage characteristics of the highly porous Bi2223 superconductor measured at different temperatures in magnetic fields of 20 and 80 Oe.

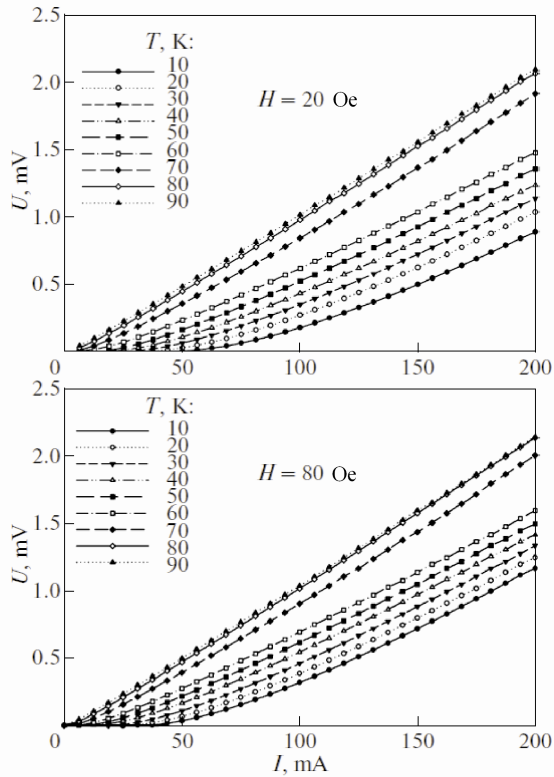

**Fig. 2.** Current–voltage characteristics of the high-temperature superconductor measured at different temperatures.

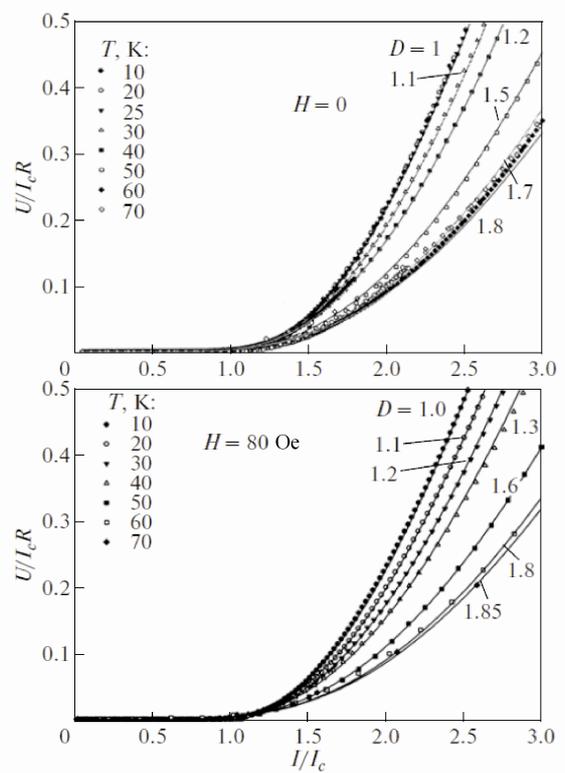

**Fig. 3.** Current–voltage characteristic normalized to the value of the critical current at different temperatures.

3. DISCUSSION OF THE RESULTS

The fractal boundaries of superconducting clusters, which, in the aforementioned sense, can be considered as both small scale and large scale defects, provide an effective pinning of the magnetic flux. The pinning by fractal boundaries between a high-temperature superconductor and a normal phase was considered in the works by Kuzmin [2,3]. The model of a superconductor with fractal clusters is suitable for the description of the transport properties of the high-temperature superconductor $Bi_{1.8}Pb_{0.3}Sr_{1.9}Ca_2Cu_3O_x$ with a low density [4]. This paper presents the results of a detailed investigation of a porous superconductor within the framework of the fractal cluster model.

For a comparison between the experimental current–voltage characteristics and theoretical curves obtained in the theory of fractal clusters, we have used the formula given in [3]:

$$u = r_f \left[ i \exp\left(-Ci^{-2/D}\right) - C^{D/2} \Gamma\left(1 - \frac{D}{2}, Ci^{-2/D}\right) \right]$$

where $D$ is the fractal dimension, $u$ is the dimensionless voltage, $r_f$ is the dimensionless resistance to the vortex flow, $i$ is the dimensionless current obtained after the normalization to the critical current $I_c$, $C = ((2 + D)/2)^{2/D + 1}$, and $\Gamma$ is the gamma function.

The normalized current–voltage characteristics at different temperatures without a magnetic field and in magnetic field of 80 Oe are shown in Fig. 3. At temperatures in the range from 40 to 50 K, there occurs a maximum change in the resistance, the slope of the linear portion of the dependences $U(I)$. This is probably due to the maximum change in the fractal dimension of the boundary of the nonsuperconducting cluster. The fractal dimension of the cluster boundaries weakly increases with an increase in the magnetic field.

In our previous work [4], we used the method proposed in [22] and obtained the fractal dimension of the grain boundaries in the highly porous sample $D = 1.7\pm0.1$. This value is close to the fitting value of the fractal dimension of the boundaries between clusters of the superconducting and normal phases $D = 1.75$, which was used in the description of the current–voltage characteristics of highly porous high-temperature superconductors at a temperature above $T = 77.4$ K in a zero magnetic field. A comprehensive picture of the evolution of the fractal dimension as a function of the temperature and the magnetic field is shown in Fig. 4.

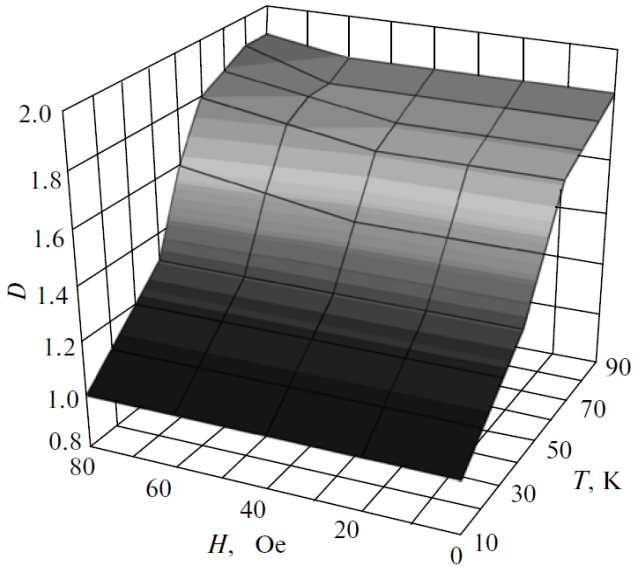

**Fig. 4.** Dependence of the fractal dimension of the normal phase boundary on the magnetic field and temperature.

In order to obtain information on the phase transitions in the vortex structure of the superconductor under investigation, we constructed the dependences $R(I)$ in logarithmic coordinates. Figure 5 presented the dependences $R(I)$ at temperatures in the range from 10 to 90 K in external magnetic fields of 20 and 80 Oe.

It can be seen from Fig. 5 that no change in the sign of the curvature from positive to negative is observed in the dependences $R(I)$ over the entire temperature range. Thus, in the measured intervals of magnetic fields and temperatures, we did not reveal phase transitions of the vortex structures. It should be noted that, in different magnetic fields, the slope of the linear portions of the curves $R(I)$ at the corresponding temperatures varies only insignificantly, and the character of the changes remains identical both in the case of the absence of an external magnetic field and in the maximum field. The fractal dimension weakly depends on the magnetic field in the range 0–80 Oe.

In the temperature range from 60 to 70 K, there is a "gap" in the series of dependences $R(I)$. At the same temperatures, the character of the temperature dependence of the differential resistance changes abruptly (Fig. 6). A similar result was obtained from theoretical calculations carried out using the Monte Carlo method [23]. The results of the calculations of the heat capacity and the current–voltage characteristics have revealed a jump in the same temperature range, which was attributed by the authors to the step-by-step melting of the lattice of Abrikosov vortices ending at the pinning centers. The similarity of the obtained current–voltage characteristics and the closeness of the temperature range, in which the voltage drop undergoes a jump, indicate a similar behavior of the highly porous superconductors and the model underlying the performed calculation. It is possible that the proposed mechanism, which indicates the occurrence of a phase transition in the temperature range from 60 to 70 K, takes place in the highly porous high-temperature superconductor. In this case, the pinning potential can be estimated as $U_p \approx 0.05$ eV.

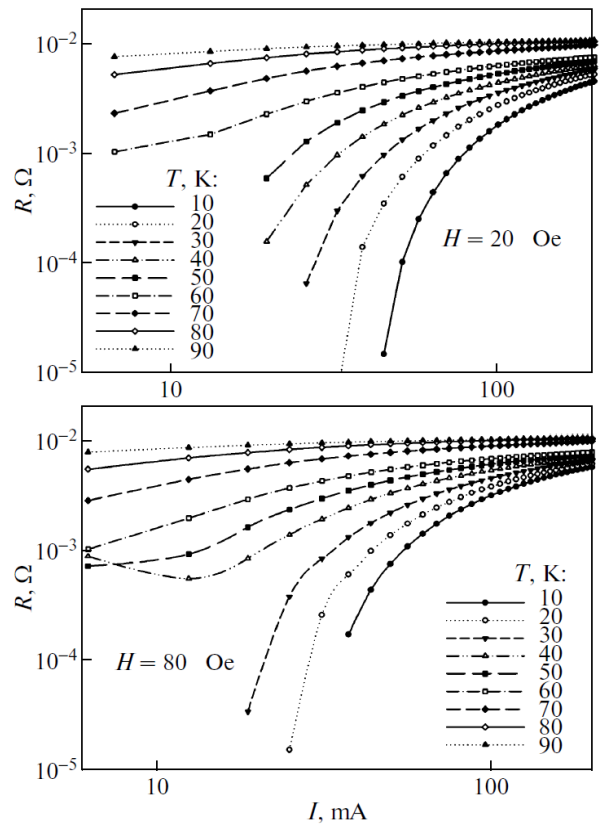

**Fig. 5.** Dependences $R(I)$ at different temperatures.

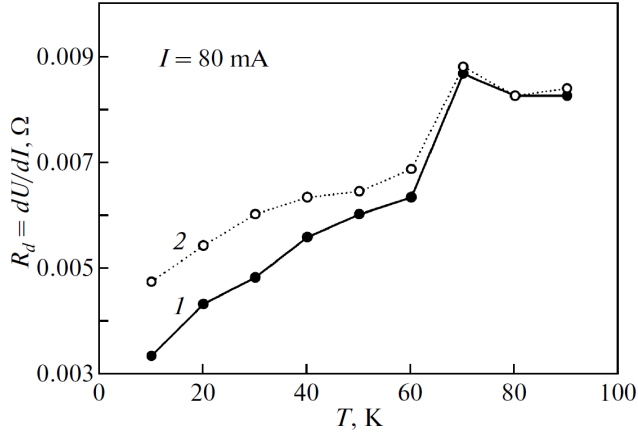

**Fig. 6.** Dependences $R_d(I)$ measured at the current $I$ = 80 mA in magnetic fields of (*1*) 20 and (*2*) 80 Oe.

With an increase in the temperature, when the energy of thermal fluctuations $k_B T$ becomes comparable to the energy of pinning $U$, the depinning of the vortices from the pinning centers begins to occur. According to the Kim–Anderson theory [24], under these conditions, there appears a finite electrical resistance $\rho = \rho_{FF}\exp(-U/k_B T)$, where $\rho_{FF}$ is the resistance to the vortex flow. The "gap" observed at temperatures in the range from 60 to 70 K can be explained by the action of two opposing effects: (i) the depinning of vortices from the pinning centers due to thermal fluctuations and (ii) the increase in the pinning due to the increase in the dimension of the fractal boundaries. At temperatures below 60 K, the increase in the dimension of the fractal boundaries partially compensates for the increase in the resistance due to thermal fluctuations. When the maximum value of the fractal dimension is achieved, this compensation is terminated and an abrupt increase in the electrical resistance is observed. An alternative explanation for the intensive increase in the electrical resistance at temperatures in the range from 60 to 70 K lies in the fact that, in this range, the core of the Abrikosov vortex becomes larger than the maximum selfsimilarity scale of the distribution of the pinning centers.

The proposed mechanisms of coupling between the fractal dimension and the specific features of the pinning in highly porous high-temperature superconductors can work fairly well both individually and collectively.

## 4. CONCLUSIONS

Thus, the experimental current–voltage characteristics of the highly porous high-temperature superconductor Bi-2223 have been measured over wide ranges of temperature (10–90 K) and have been analyzed within the framework of the theory of phase transformations of vortex matter [21] and the theory of a fractal boundary of the normal phase cluster in the superconducting matrix [2, 3]. The absence of a change in the sign of the curvature of the dependences $R(I)$ with an increase in the temperature indicates, according to [21], that no phase transition of the vortex structure occurs. A noticeable increase in the electrical resistance has been observed in the temperature range 60–70 K, which, as we assume, is associated with the change in the fractal dimension of the boundary of the normal phase cluster growing inward the crystallites of the highly porous high-temperature superconductor [2, 3]. As this temperature interval is achieved, the core of the Abrikosov vortex becomes larger than the size of the maximum self-similarity scale of the fractal interface and the effective pinning drastically decreases. In this case, vortex matter undergoes a step-by-step melting, which is in agreement with the results reported in [23]. A possible mechanism of this phenomenon is provided by a partial compensation of the increase in the electrical resistance due to thermal fluctuations at the expense of the increase in their dimension at temperatures in the range from 50 to 60 K. When the maximum value of the fractal dimension is achieved, this compensation is reduced and an abrupt increase in the electrical resistance is observed.

When this paper was already prepared for publication, an article by Vasyutin [25] appeared, in which it was found that the fractal dimension of the channels for vortex transport in YBCO increases with increasing magnetic field.